\documentclass[lettersize,journal]{IEEEtran}
\usepackage{amsmath,amsfonts}
\usepackage{algorithmic}
\usepackage{array}
\usepackage[caption=false,font=footnotesize,labelfont=rm,textfont=rm]{subfig}
\usepackage{textcomp}
\usepackage{stfloats}
\usepackage{url}
\usepackage{cite}
\usepackage{verbatim}
\usepackage{bm}
\usepackage{graphicx}
\usepackage{amssymb}
\usepackage{amsmath,amsfonts}
\usepackage{extarrows}
\usepackage{amsmath,amsthm,amssymb,amsfonts}
\usepackage{algorithm}
\usepackage{algorithmic}
\usepackage{hyperref} 
\hypersetup{
	hidelinks,
	colorlinks=true,
	linkcolor=red,
	citecolor=blue,
	urlcolor = black}
\def\BibTeX{{\rm B\kern-.05em{\sc i\kern-.025em b}\kern-.08em
		T\kern-.1667em\lower.7ex\hbox{E}\kern-.125emX}}
\usepackage{balance}
\everymath{\displaystyle}

\begin{document}

\title{DRL-Based Secure Transmission for Rotatable Antenna-Enabled Low-Altitude ISAC Systems}

\author{Chuan Liu, Hongyi Bian,  Wei Gao, Qi Zhang, Yu Yao, Liang Yang, and Feng Shu
	\thanks{Chuan Liu is with the School of Electronic Information and Communication, Huazhong University of Science and Technology, Wuhan 430074, China, and also with the Department of Information and Communication, China Electric Power Research Institute, Beijing 100192, China (e-mail: liu chuan@hust.edu.cn).}
	\thanks{Hongyi~Bian,  Wei Gao , Qi Zhang and Yu~Yao are with the School of Information and Communication Engineering, Hainan University, Haikou, 570228, China (e-mail: hongyibian@hainanu.edu.cn; gaowei@hainanu.edu.cn; hdzhangqi0509@163.com; shell8696@hotmail.com).}
	
	\thanks{Liang Yang is with the College of Computer Science and Electronic Engineering, Hunan University, Changsha 410082, China, and also with the chool of Communication and Electronic Engineering, Jishou University, jishou, Hunan 416000, China (e-mail: liangy@hnu.edu.cn).}
    
	\thanks{Feng~Shu is with the School of Information and Communication Engineering, Hainan University, Haikou, 570228, China, and also with the School of Electronic and Optical Engineering, Nanjing University of Science and Technology, Nanjing, 210094, China (e-mail: shufeng0101@163.com).}}

%
%

\maketitle

\begin{abstract}

The development of the low-altitude economy has driven innovation in intelligent antenna systems within ISAC systems. In this paper, we investigate a  Rotatable Antenna (RA)-enabled  low-altitude integrated sensing and communication (ISAC) system. In practical terms, the RA array can flexibly adjust the three-dimensional (3D) beam direction of each antenna to enhance array directional gain, thereby improving the communication security of legitimate mobile users against potential eavesdropping risks from the unmanned aerial vehicle (UAV). Our objective is to maximize the minimum secrecy rate (SR) by jointly optimizing transmit beamforming matrix, transmit and receive RAs' pointing matrices. To this end, an multi-agent proximal policy optimization with three  improvement  mechanisms (MAPPO-T) algorithm is proposed to cope with the issue of complex multi-agent collaborative decision-making problem. Simulation results show that the introduction of RAs can effectively improve SR performance compared to the traditional fixed orientation antenna (FOA)-based system. In addition, the proposed MAPPO-T algorithm validate the superiority compared to the standard MAPPO algorithm.
  
\end{abstract}

\begin{IEEEkeywords}
Rotatable Antenna; integrated sensing and communication (ISAC); multi-agent proximal policy optimization (MAPPO).
\end{IEEEkeywords}

\section{Introduction}
\subsection{Background}
With the development of the low-altitude economy, low-altitude flight activities are becoming increasingly normalized and dense, posing a severe challenge to existing wireless networks. The application scenarios in low-altitude environments require the coverage model of wireless networks to evolve from traditional ground coverage to a three-dimensional (3D) integrated coverage that encompasses both air and ground. Therefore, constructing a low-altitude communication network that meets the needs of low-altitude integrated sensing and communication (ISAC) \cite{Liu2022, Mao2025, Yao2025}, achieving air-ground coverage, stable connections, and manageable control, has become a key technological foundation for supporting the large-scale development of the low-altitude industry \cite{Yao2026, Jinbing2022}. However, existing antenna systems that utilize fixed antenna architectures typically have a limited range of radiation angle adjustment, making it difficult to achieve continuous sensing and beam alignment for low-altitude unmanned aerial vehicles (UAVs).

The inherent limitations of fixed antennas in lacking three-dimensional spatial degree of freedom (DoF) have led to the emergence of rotatable antennas (RAs) based on non-fixed architectures as a groundbreaking alternative \cite{Tan2026, Shao2026}. Notably, in low-altitude wireless networks, the ability to control the orientation of the RA's line of sight holds particularly significant potential value. The reason is the antenna beams of traditional ground base stations (BSs) are typically optimized for ground users, resulting in limited vertical coverage that struggles to effectively support the ISAC needs of low-altitude aircraft \cite{Zhang2026a}. Moreover, RAs can flexibly adjust the orientation of the main lobe of radiation through mechanical or electronic means, enabling precise focusing of signal energy on low-altitude targets, thereby expanding the coverage range in the vertical dimension and enhancing link quality \cite{Li2026}. Furthermore, the beam scanning characteristics of RAs facilitate continuous tracking and high-precision sensing of high-speed moving targets, providing crucial technical support for low-altitude airspace management. 

The unique beam tracking and scanning capabilities of RAs can enhance sensing accuracy and coverage in ISAC systems, which improve signal quality and effectively suppress interference by leveraging flexible beamforming and spatial multiplexing gains \cite{Wang2025a, Feng2026}. More interestingly, RAs can adaptively scan spatial sectors, enhancing sensing coverage and resolution without the need for complex mechanical platforms. Furthermore, by independently controlling the orientation of the visual axes of different antennas in the array, RAs can collaboratively optimize the angular resolution of the sensing area and the link quality for communication users, thereby improving coverage efficiency and robustness of environmental perception in complex settings. 

In UAV communication and integrated air-ground-space networks, ubiquitous connectivity is a key capability that supports the transition of wireless networks from single spatial coverage to seamless global coverage \cite{Chen2026, Yao2026a}. RAs can flexibly respond to the rapid trajectory changes of aerial targets by dynamically adjusting the orientation of their visual axes, thereby extending the 3D spatial coverage capability of BSs \cite{Zhang2026b}. The core of RAs advantage lies in addressing the limitations of traditional downward-tilted antennas in supporting aerial nodes, which can efficiently cover airspace relying on ground BSs, as well as be deployed on aerial platforms to ensure continuous communication for moving targets through dynamic beam tracking technology.

With the deepening of research, intelligent steering antenna technology has gradually been applied in cutting-edge fields such as secure communication \cite{Li2026a}. The channel characteristics of traditional fixed antenna arrays exhibit static replicability, making them susceptible to precise matching and interception by eavesdroppers \cite{Wang2025}. In contrast, the time-varying characteristics introduced by the rotating structure effectively overcome this inherent flaw, making it difficult for eavesdroppers to track the signal transmission patterns, thus preventing effective interception \cite{Shu2021}. The rotation of RAs synchronously drives the real-time switching of the radiation pattern and spatial beam direction. When the legitimate communication user is located in the preset desired transmission direction, RAs can continuously receive high-gain main lobe signals, ensuring the quality of link transmission; whereas eavesdropping nodes located in undesired directions can only capture low-gain sidelobe signals or deeply faded signals, thus failing to obtain effective communication information.

\subsection{Overview of Related Work}

The application of RAs requires extensive channel modeling and channel estimation experiments to lay the groundwork for wireless transmission research.  Therefore, numerous researchers are dedicated to analyzing the challenges that RAs pose for acquiring channel state information (CSI) in systems \cite{Xiong2026, Zheng2025a, Zheng2026, Wu2025, Xiong2025, Wang2026}.
Xiong et al. \cite{Xiong2026} investigated the fundamental principles of a novel intelligent RA and its system performance advantages, while discussing the main design issues of integrated sensing and communication systems based on RAs, and proposed corresponding solutions. Zheng et al. \cite{Zheng2025a} systematically exploreed the technical principles and application scenarios of intelligent steering antennas, revealing their flexible adaptability in communication and sensing tasks. Zheng and Wu et al. \cite{Zheng2026, Wu2025} established theoretical models for intelligent steering antenna-enabled wireless communication systems, covering system modeling, channel characterization, and performance analysis.  Reference \cite{Xiong2025} further proposed an efficient channel estimation algorithm suitable for intelligent steering antenna systems, effectively improving channel estimation accuracy and system stability. To overcome the increased cost of CSI estimation, authors in \cite{Wang2026} further proposed a two-timescale channel modeling and optimization framework for six-dimensional (6D) RA arrays in cell-free networks, separating large-timescale 6D RA arrays control from small-timescale signal processing. 

As one of the most extensively researched technological paradigms currently, ISAC has also seen rotating antennas widely adopted by numerous researchers to advance the core technologies in this field.
Zhou et al. \cite{Zhou2025} improved communication and sensing performance by introducing RAs to increase DoF. In order to reduce the operating costs of hardware, authors in \cite{Qu2025} investigated multiple-BS enabled by RAs to enhance the efficient utilization of spatial resources in ISAC systems. Sun et al. \cite{Sun2025} theoretically analyzed the gain mechanism of array rotation for near-field ISAC, filling the theoretical gap of rotating antennas in near-field integrated sensing and communication. Authors in \cite{Wang2026a} proposed a low-cost ISAC solution combining RAs with sub-connected hybrid beamforming to achieve cost-effective integrated sensing and communication performance. Zhang et al. \cite{Zhang2026} pioneered a near-field ISAC system framework that combines rotatable and movable antennas, dynamically adjusting antenna positions and 3D rotation to simultaneously enhance communication rates and sensing accuracy. In addition,  Reference \cite{Pei2024, Pei2025}  employed a rotatable half-wavelength antenna for UAV-assisted ISAC, providing a feasible solution for lightweight deployment of ISAC in low-altitude scenarios.

However, security issues still pose certain challenges for researchers in the field of RA studies, which has consequently attracted in-depth exploration from a wide range of scholars. Dai et al. \cite{Dai2026} first applied RA arrays to covert communication, effectively suppressing the detection probability of the monitor (Willie) through spatial domain directional control. Authors in \cite{Dai2025} combined RAs with MIMO physical layer security, enhancing system security performance through rotational directional gain reshaping. Jiang et al. \cite{Jiang2025} proved the critical theory that the average secrecy capacity exhibits quasi-concavity with respect to the rotation adjustment factor in non-real-time adjustable RA security communication scenarios, and derived the important conclusion that the secrecy outage probability is independent of the rotation angle at high signal-to-noise ratios.

\subsection{Motivations and Contributions}
Based on the aforementioned discussion, we study the secure transmission for RA-enabled low-altitude ISAC systems. Different from traditional fixed location users, the consideration for mobile users in the article is more in line with practical scenarios.  In addition, most traditional methods are difficult to cope with mobile users and target scenarios, which failing to derive valid solutions to adapt the dynamic environment. To this end, we are committed to design a innovative deep reinforcement learning (DRL) algorithm to effectively solve collaborative decision-making problems in multi-agent environments. The main contributions are listed as follows:
\begin{itemize}
	\item{}In this paper, we propose a novel RA-enabled secure transmission for low-altitude ISAC systems, where the rotation angle of RA is adjusted dynamically at BS to enhance the system's secure transmission and target sensing capabilities. In addition, the mobility of all users places the system in a dynamic environment subject to temporal changes. Thus, the max-min secrecy rate (SR) problem is formulated by jointly optimizing the transmit beamforming of BS 	and the pointing matrices of RAs, respectively.
	\item{} To address the high dimensional action problem in complex dynamic environment, we introduced the multi-agent proximal policy optimization with three  improvement  mechanisms (MAPPO-T) algorithm to tackle multi-agent collaborative decision-making. Specifically, to address the potential limitations of the standard MAPPO algorithm, we have incorporated a noise-based approach to mitigate overfitting and introduced an improvement based on recurrent neural networks (RNN) to handle time-series states. Additionally, we have implemented the Preserving Outputs Precisely while Adaptively Rescaling Targets (PopArt) method to address training instability issues.
	\item{} The comprehensive simulation results demonstrate the advantages  of the RA-enable ISAC systems, which can effectively enhance both secure transmission rate and sensing performance without requiring additional antennas.  Furthermore, compared to the standard MAPPO algorithm and other benchmark schemes, the proposed MAPPO-T algorithm achieves significant improvements in secure transmission performance. 
	
\end{itemize}

\subsection{Organization and Notation}
The remainder of the paper is organized as follows.  In Section \ref{System Model}, we present the RA-enabled low-altitude ISAC system and formulate the max-min SR optimization problem. Section \ref{MAPPO}  provides the proposed MAPPO-T algorithm to solve the thorny issue. The performance evaluation is presented in Section \ref{Simulation Results}. Lastly, Section \ref{Conclusion} provides the concluding remarks of this paper.

$\mathrm{Notation}$: In this paper, scalars, vectors, and matrices are signified by standard fonts, bold lowercase letters, and bold uppercase letters, respectively. The conjugate, transpose, and Hermitian transpose operations are denoted by $(\cdot)^{\ast}$, and $(\cdot)^T$, $(\cdot)^H$, respectively. $|\cdot|$, $\|\cdot\|$, and $\|\cdot\|_F$ stand for the norm, $l_2$-norm, and Frobenius-norm, respectively.  $\mathbf{I}_N$ corresponds to a $N$-dimensional identity matrix and $\mathcal{CN}(0,\sigma^2)$ refers to the complex normal distribution with zero-mean and variance $\sigma^2$. In addition, $\mathbb{E}\{\cdot\}$ signifies the expectation operator and $[\cdot]^+$  is a value greater than or equal to zero.

\section{System Model}
\label{System Model}

\begin{figure}[!t]
	\centering
	\includegraphics[width=0.48\textwidth]{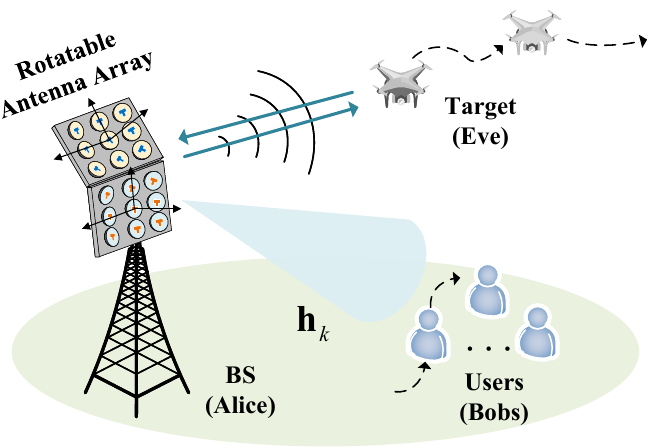}\\
	\caption{System model for active RIS-assisted secure transmission with RA arrays.}\label{f1}
\end{figure}
As illustrated in Fig. \ref{f1}, $M = M_x M_z$ transmit rotatable uniform planar arrays (UPA)  confidential information to $K$ legitimate users (Bobs) via  the line-of-sight (LoS) path, where $M_x$ and $M_z$ represent the number of RAs  along the horizontal and vertical directions, respectively. The UAV (Eve) to be detected eavesdrops on useful information from $K$ mobile users while moving irregularly at a certain altitude in the air. 
Similarly, the receiving UPA is equipped with $N = N_x N_z$ antennas connected to a single radio frequency chain, where $N_x$ and $N_z$ denote the number of RAs along the horizontal and vertical directions, respectively. The distance between adjacent RAs is denoted by $d$, and thus the transmit and receiving UPA size can be mathematically described as $A = M_x d \times M_y d$ and $A = N_x d \times N_y d$, respectively. 
Assuming the BS is equipped with directional antennas, while Bob and Eve are each equipped with a single isotropic antenna. Let $\mathcal{N}_x=\{1,\ldots,N_x-1\}$ and $\mathcal{N}_z=\{1,\ldots,N_z-1\}$ denote the index sets of RAs along the the horizontal and vertical directions within the transmit rotatable UPA, respectively. Similarly, let $\mathcal{M}_x=\{1,\ldots,M_x-1\}$ and $\mathcal{M}_z=\{1,\ldots,M_z-1\}$ denote the index sets of RAs along the horizontal and vertical directions within the receive rotatable UPA, respectively.

In the three-dimensional coordinate system, the $m$-th transmitting RA located at the $m_x$-th row and $m_z$-th column is used as the reference element, and its position can be denoted as
\begin{equation}
	\mathbf{a}_m\triangleq\mathbf{a}_{m_x,m_z}\triangleq\mathbf{a}_{(m_x-1)M_z+m_z}=[(m_x-1)d,0,(m_z-1)d]^T,
\end{equation}
where $m_x \in \mathcal{M}_x$ and $m_z \in \mathcal{M}_z$. The indices $M_x$ and $M_z$ can be mapped to $m$. Similarly, the position of  the receiving RA located at the $n_x$-th row and $n_z$-th column  can be denoted as
\begin{equation}
	\mathbf{b}_n\triangleq\mathbf{b}_{n_x,n_z}\triangleq\mathbf{b}_{(n_x-1)N_z+N_z}=[(n_x-1)d,0,(n_z-1)d]^T,
\end{equation}
where $m_x \in \mathcal{M}_x$, $m_z \in \mathcal{M}_z$. Let $r_{k}[l]$ represents the distance between the center of the transmit RA array and Bob $k$. Thus, the position of $k$-th Bob  can be denoted by
\begin{equation}
	\begin{aligned}
	\mathbf{u}_{k}[l]&=\left[u_{k,x}[l], u_{k,y}[l], u_{k,z}[l]\right]^T\\
	&=\left[r_{k}[l]\Theta_k[l],r_{k}[l]\Lambda_k[l],r_{k}[l]\Xi_k[l]\right]^T,
	\end{aligned}
\end{equation}
with $\Theta_k[l]\triangleq\sin\psi_k[l]\sin\phi_k[l]$, $\Lambda_k[l]\triangleq\cos\psi_k[l]$,  $\Xi_k[l]\triangleq\sin\psi_k[l]\cos\phi_k[l]$, where $\psi_{k}[l]\in[0,\pi]$ and $\phi_{k}[l]\in[-{\pi}/{2},{\pi}/{2}]$ represent the zenith and azimuth angles from RA array to Bob $k$, respectively. The velocity of $k$-th mobile user  is $\mathbf{v}_{k}[l]=[v_{k,x}[l],v_{k,y}[l],0]^T$. 
Similarly, the position of Eve at time slot $t$ can be denoted by
\begin{equation}
	\begin{aligned}
		\mathbf{u}_{\mathrm{e}}[l]=[e_x[l],e_y[l],e_z[l]]^T,
	\end{aligned}
\end{equation}
with velocity $\mathbf{v}_{e}[l]=[v_{e,x}[l],v_{e,y}[l],v_{e,z}[l]]^T$, where $e_x[l]\triangleq r_{\mathrm{e}}[l]\sin\psi_{\mathrm{e}}[l]\sin\phi_{\mathrm{e}}[l]$, $e_y[l]\triangleq r_{\mathrm{e}}[l]\cos\psi_{\mathrm{e}}[l]$, and $e_z[l]\triangleq r_{\mathrm{e}}[l]\sin\psi_{\mathrm{e}}[l]\cos\phi_{\mathrm{e}}[l]$.
$r_e[l]$ represents the distance between the center of the transmit RA array and Eve at the time slot $l$. $\psi_{\mathrm{e}}[l]\in\Delta_{\psi}\in[0,\pi]$ and $\phi_{\mathrm{e}}[l]\in\Delta_{\phi}\in[-{\pi}/{2},{\pi}/{2}]$ represent the zenith and azimuth angles from RA array to Willie at time slot $l$, where $\Delta_{\psi}$ and $\Delta_{\phi}$ denote the estimation errors for the zenith and azimuth angles, respectively. Accordingly, the distance from  $k$-th Bob to $m$-th RA  is expressed as
\begin{equation}
	\begin{aligned}
		&r_{k,m}[l]=\|\mathbf{u}_{k}[l]-\mathbf{a}_{m}\|=r_{k}[l] \times\\
		&\sqrt{1-2m_{x}\delta_{k}[l]\Theta_k[l]-2m_{z}\delta_{k}[l]\Lambda_k[l]+(m_{x}^{2}+m_{z}^{2})\delta_{k}^{2}[l]},
	\end{aligned}
\end{equation}
\begin{equation}
	\begin{aligned}
		&r_{e,m}[l]=\|\mathbf{u}_{e}[l]-\mathbf{a}_{m}\|=r_{e}[l] \times\\
		&\sqrt{1-2m_{x}\delta_{e}[l]\Theta_e[l]-2m_{z}\delta_{e}[l]\Lambda_e[l]+(m_{x}^{2}+m_{z}^{2})\delta_{k}^{2}[l]},
	\end{aligned}
\end{equation}
where $\delta_{k}[l]\triangleq\frac{d}{r_{\mathrm{t},k}[l]}$ and $\delta_{\mathrm{e}}[l]\triangleq\frac{d}{r_{\mathrm{e}}[l]}$.
\begin{figure}[!t]
	\centering
	\subfloat[Transmit rotational angle of the $m$-th RA.]{%
		\includegraphics[width=0.45\linewidth]{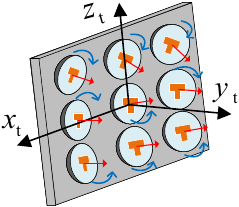}%
		\label{fig:global_data}%
	}\hfill
	\subfloat[Transmit rotational angle of the $n$-th RA.]{%
		\includegraphics[width=0.45\linewidth]{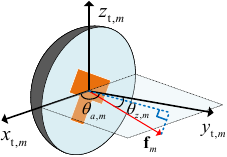}%
		\label{fig:long_term_data}%
	}
	\caption{Illustration of transmit rotational angles and directional gain pattern.} 
	\label{rotational angle} 
\end{figure}
As shown in Fig. \ref{rotational angle},  the pointing vector of the $m$-th and $n$-th of the transmit and receiving UPA can be respectively denoted as 
\begin{equation}
	\begin{aligned}
		&\mathbf{f}_{m}[l]=\\
		&\left[\sin\theta_{\mathrm{z},m}[l]\cos\theta_{\mathrm{a},m}[l],\sin\theta_{\mathrm{z},m}[l]\sin\theta_{\mathrm{a},m}[l],\cos\theta_{\mathrm{z},m}[l]\right]^T,
	\end{aligned}
\end{equation}
\begin{equation}
	\begin{aligned}
	\mathbf{t}_{n}[l]=\left[\sin\theta_{\mathrm{z},n}[l]\cos\theta_{\mathrm{a},n}[l],\sin\theta_{\mathrm{z},n}\sin\theta_{\mathrm{a},n}[l],\cos\theta_{\mathrm{z},n}[l]\right]^T,
	\end{aligned}
\end{equation}
where $\theta_{\mathrm{z},m}[l]$, $\theta_{\mathrm{a},m}[l]$, $\theta_{\mathrm{z},n}[l]$ and $\theta_{\mathrm{a},n}[l]$ represent the zenith and azimuth angles of transmit and receive RAs, respectively, and $\|\mathbf{f}_{m}[l]\|=\|\mathbf{f}_{n}[l]\|=1$. Note that all RAs within the same UPA share an pointing vector. In practice, the boresight direction of RAs can be controlled jointly or separately via mechanical and electronic methods, subject to limited rotation precision and range. The rotation range of zenith angles can be denoted by
\begin{equation}\label{jiaodu1}
	\begin{array}{cc}
		0\leq\theta_{\mathrm{z},m}[l]\leq\theta_{\mathrm{max}},&\forall m,
	\end{array}
\end{equation}
\begin{equation}\label{jiaodu2}
	\begin{array}{cc}
		0\leq\theta_{\mathrm{z},n}[l]\leq\theta_{\mathrm{max}},&\forall n,
	\end{array}
\end{equation}
where $\theta_{\mathrm{max}}\in[0,\tfrac{\pi}{2}]$ denotes the maximum zenith angle. 
The minimal adjustable steps for the zenith and azimuth angles are denoted by $\theta_{z,\mathrm{min}}$ and $\theta_{a,\mathrm{min}}$, respectively. 
The sets of all selectable angles for the zenith and azimuth angles of the $m$-th and $n$-th RA can be respectively denoted as 
\begin{subequations}
	\begin{alignat}{2}
		\Gamma_{\theta_{\mathrm{z},m}}&=\{0,\theta_{\mathrm{z},\mathrm{min}},\ldots,(I_{\mathrm{z},m}-1)\theta_{\mathrm{z},\mathrm{min}}\},\\
		\Gamma_{\theta_{\mathrm{a},m}}&=\{0,\theta_{\mathrm{a},\mathrm{min}},\ldots,(I_{\mathrm{a},m}-1)\theta_{\mathrm{a},\mathrm{min}}\},\\
		\Gamma_{\theta_{\mathrm{z},n}}&=\{0,\theta_{\mathrm{z},\mathrm{min}},\ldots,(I_{\mathrm{z},n}-1)\theta_{\mathrm{z},\mathrm{min}}\}, \\
		\Gamma_{\theta_{\mathrm{a},n}}&=\{0,\theta_{\mathrm{a},\mathrm{min}},\ldots,(I_{\mathrm{a},n}-1)\theta_{\mathrm{a},\mathrm{min}}\},
	\end{alignat}
\end{subequations}
where $I_{\mathrm{z},m}$, $I_{\mathrm{a},m}$, $I_{\mathrm{z},n}$ and $I_{\mathrm{a},n}$ represent the numbers of adjustable zenith and azimuth angles of $m$-th and $n$-th RA, respectively. Additionally, $I_m = I_{\mathrm{z},m} I_{\mathrm{a},m}$ and $I_n = I_{\mathrm{z},n} I_{\mathrm{a},n}$ denote the total number of pointing vectors of the $m$-th and $n$-th RA. As described in reference \cite{Shao2025}, to avoid mutual signal reflection between the transmitting and receiving RAs, they must satisfy specific rotational constraints. Our work addresses the issue by maintaining a large separation between the transmitting RA and receiving RA, thus ensuring the absence of mutual signal reflection.

\subsection{Channel Model}
In this article, we investigate downlink multiuser transmission with a quasi-static channel model and assume that Bobs and the sensing target are located in the near-field region of the BS.
Considering multipath channels, according to \cite{Zheng2026},  the line-of-sight (LoS) channel component $h_{k,m}^{\mathrm{LoS}}$ and the non-LoS (NLoS) channel component $h_{k,m}^{\mathrm{NLoS}}$ from the $m$-th RA to Bob $k$ can be expressed as
\begin{equation}
	\begin{aligned}
	h_{k,m}^{\mathrm{LoS}}[l] = \sqrt{\tfrac{A}{4\pi r_{k,m}^{2}[l]}G_{0}\cos^{2p}(\omega_{k,m}[l])}e^{-j2\frac{\pi}{\lambda}r_{k,m}[l]},
	\end{aligned}
\end{equation}
\begin{equation}
	\begin{aligned}
		h_{k,m}^{\mathrm{NLoS}}[l]&=\sum_{q=1}^Q\!\sqrt{\tfrac{\sigma_qA}{4\pi {r}_{q,m}^{2}{r}_{k,q}}G_{0}\cos^{2p}({\omega}_{q,m})}\\
		&\times e^{-j2\frac{\pi}{\lambda}({r}_{q,m}+{r}_{k,q})+j\tau_q},
	\end{aligned}
\end{equation}
where  $p$ denotes the directivity factor of RA, $G_{0}=2(2p+1)$ denotes the maximum gain in the boresight direction, and $\cos(\omega_{k,m}[l])\triangleq \mathbf{f}_{m}^T[l] \mathbf{u}_{k,m}[l]$ denotes the projection between $\mathbf{f}_{m}[l]$ and $\mathbf{u}_{k,m}[l]$ with $\mathbf{u}_{k,m}[l]\triangleq\frac{\mathbf{u}_k[l]-\mathbf{a}_m}{\|\mathbf{u}_k[l]-\mathbf{a}_m\|}$ denoting the
direction vector from RA $m$ to Bob $k$. $\mathbf{c}_{q}[l] \in\mathbb{R}^{3\times 1}$ is the position of the $q$-th scatterer cluster, and $\tau_q$ denotes the phase shift corresponding to the $q$-th scatterer cluster.  
$\omega_{k,m}[l]$ denotes the angle between the boresight direction and Bob $k$. $\cos(\omega_{q,m}[l])\triangleq \mathbf{f}_{m}^T[l] \mathbf{c}_{q,m}[l]$ denotes the projection between $\mathbf{f}_{m}[l]$ and $\mathbf{c}_{q,m}[l]$ with $\mathbf{c}_{q,m}[l]\triangleq\frac{\mathbf{c}_q[l]-\mathbf{a}_m}{\|\mathbf{c}_q[l]-\mathbf{a}_m\|}$ being the
direction vector from RA $m$ to scatterer cluster $q$,  where $\omega_{q,m}$ denotes the angle between the boresight direction and the $q$-th scatterer cluster, respectively.
 ${r}_{q,m}[l]=\|\mathbf{c}_{q}[l]-\mathbf{a}_{m}\|$ and ${r}_{k,q}[l]=\|\mathbf{u}_{k}[l]-\mathbf{c}_{q}[l]\|$  represent the distances from $q$-th ($q = 1, 2, \ldots, Q$) scatterer cluster to $m$-th RA, and from $k$-th Bob to $q$-th scatterer cluster, respectively. 
Similarly,  the LoS channel component $h_{\mathrm{e},m}^{\mathrm{LoS}}[l]$ and the NLoS channel component $h_{\mathrm{e},m}^{\mathrm{NLoS}}[l]$ from the $m$-th RA to Eve can be expressed as
\begin{equation}
	\begin{aligned}
	h_{\mathrm{e},m}^{\mathrm{LoS}}[l] = \sqrt{\tfrac{A}{4\pi r_{\mathrm{e},m}^{2}[l]}G_{0}\cos^{2p}(\omega_{\mathrm{e},m}[l])}e^{-j2\frac{\pi}{\lambda}r_{\mathrm{e},m}[l]},\\
	\end{aligned}
\end{equation}
\begin{equation}
	\begin{aligned}
		&h_{\mathrm{e},m}^{\mathrm{NLoS}}[l]=\sum_{q=1}^Q\sqrt{\tfrac{A}{4\pi {r}_{q,m}^{2}[l]{r}_{\mathrm{e},q}^{2}[l]}G_{0}\cos^{2p}({\omega}_{q,m}[l])} \\
		&\times e^{-j2\frac{\pi}{\lambda}({r}_{q,m}[l]+{r}_{\mathrm{e},q}[l])+j{\tau}_q},
	\end{aligned}
\end{equation}
where $r_{\mathrm{e},m}[l]=\|\mathbf{u}_{\mathrm{e}}[l]-\mathbf{a}_{m}\|$ and ${r}_{\mathrm{e},q}[l]=\|\mathbf{u}_{\mathrm{e}}[l]-\mathbf{c}_{q}\|$ represent the distances from $m$-th RA  to Eve and from $q$-th scatterer cluster to Eve, respectively. $\omega_{\mathrm{e},n}$ denotes the angles between the boresight direction and Eve. Then, the direct channel $\mathbf{h}_{k}\in\mathbb{C}^{M\times 1}$ and $\mathbf{h}_{\mathrm{e}}\in\mathbb{C}^{M\times 1}$ from BS to Bob $k$ and Eve can be separately denoted as
\begin{equation}
	\begin{aligned}
	\hat{\mathbf{h}}_{k}[l]&=\hat{\mathbf{h}}_{k}^{\mathrm{LoS}}[l]+\hat{\mathbf{h}}_{k}^{\mathrm{NLoS}}[l]\\
	&=[h_{k,1}^{\mathrm{LoS}}[l],h_{k,2}^{\mathrm{LoS}}[l],\ldots,h_{k,M}^{\mathrm{LoS}}[l]]^T\\
	&+[h_{k,1}^{\mathrm{NLoS}}[l],h_{k,2}^{\mathrm{NLoS}}[l],\ldots,h_{k,M}^{\mathrm{NLoS}}[l]]^T,
	\end{aligned}
\end{equation}
\begin{equation}
	\begin{aligned}
		\hat{\mathbf{h}}_{\mathrm{e}}[l] 
		&=\mathbf{h}_{\mathrm{e}}^{\mathrm{LoS}}[l]+\hat{\mathbf{h}}_{\mathrm{e}}^{\mathrm{NLoS}}[l]\\
		&=[h_{\mathrm{e},1}^{\mathrm{LoS}}[l],h_{\mathrm{e},2}^{\mathrm{LoS}}[l],\ldots,h_{\mathrm{e},M}^{\mathrm{LoS}}[l]]^T\\
		&+[h_{\mathrm{e},1}^{\mathrm{NLoS}}[l],h_{\mathrm{e},2}^{\mathrm{NLoS}}[l],\ldots,h_{\mathrm{e},M}^{\mathrm{NLoS}}[l]]^T.
	\end{aligned}
\end{equation}
Assume that the transmitter and receiver are sufficiently separated such that their channels are independently distributed. The normalized spatial correlation matrix $\mathbf{R}_k[l]\in\mathbb{C}^{M\times 1}$ and $\mathbf{R}_e[l]\in\mathbb{C}^{M\times 1}$ at the BS for Bob $k$  and Eve are respectively modeled as \cite{Bjoernson2021}
\begin{equation}
	\begin{aligned}
		\mathbf{R}_k[l]=\frac{1}{A}\mathbb{E}\left\{\hat{\mathbf{h}}_k[l]\hat{\mathbf{h}}_k^\mathrm{H}[l]\right\},
	\end{aligned}
\end{equation}
\begin{equation}
	\begin{aligned}
		\mathbf{R}_{\mathrm{e}}[l]=\frac{1}{A}\mathbb{E}\left\{\hat{\mathbf{h}}_e[l]\hat{\mathbf{h}}_e^\mathrm{H}[l]\right\}.
	\end{aligned}
\end{equation}

Thus, the channels of the direct link between BS and Bob $k$ and Eve can be represented as
\begin{equation}
	\begin{aligned}
		\mathbf{h}_k[l]=(\mathbf{R}_k[l])^{\frac{1}{2}}{\mathbf{c}}_k[l],
	\end{aligned}
\end{equation}
\begin{equation}
	\begin{aligned}
		\mathbf{h}_{\mathrm{e}}[l]=(\mathbf{R}_{\mathrm{e}}[l])^{\frac{1}{2}}{\mathbf{d}}_{\mathrm{e}}[l],
	\end{aligned}
\end{equation}
where ${\mathbf{c}}_k[l]$ and ${\mathbf{d}}_{\mathrm{e}}[l]$ express the corresponding fast-fading vectors and follow an independent and identically distributed Gaussian distribution.

\subsection{Communication and Radar Performance Metrics}
The transmitted signal incorporating both communication and radar functions in the $l$-th is represented as follows
\begin{equation}
	\begin{aligned}
		\mathbf{x}[l]=\mathbf{W}_{c}[l]\mathbf{s}_{\mathrm{c}}[l]+\mathbf{W}_{\mathrm{r}}[l]\mathbf{s}_{\mathrm{r}}[l]=\mathbf{W}[l]{\mathbf{s}}[l],
	\end{aligned}
\end{equation}
where $\mathbf{s}_{\mathrm{c}}[l]=[s_{{\mathrm{c}},1}[l],\dots,s_{{c},K}[l]]^T\in\mathbb{C}^{K \times 1}$ denotes the communication symbols intended for $K$ users, and $\mathbf{s}_{{\mathrm{r}}}[l]=[s_{\mathrm{r},1}[l],\dots,s_{{\mathrm{r}},M}[l]]^T\in\mathbb{C}^{M \times 1}$ denotes $M$ individual radar waveforms. $\mathbf{W}_{\mathrm{c}}[l]=[\mathbf{w}_{{\mathrm{c}},1}[l],\mathbf{w}_{{\mathrm{c}},2}[l],\cdots,\mathbf{w}_{{\mathrm{c}},K}[l]]\in\mathbb{C}^{M \times K}$ and  $\mathbf{W}_{\mathrm{r}}[l]=[\mathbf{w}_{{r},1}[l],\mathbf{w}_{{r},2}[l],\cdots,\mathbf{w}_{\mathrm{r},M}[l]]\in\mathbb{C}^{M \times M}$ represent the corresponding communication and radar
beamforming matrix, respectively.  In addition, we define the combined beamforming matrix $\mathbf{W}[l]\triangleq[\mathbf{W}_{\mathrm{c}}[l], \mathbf{W}_{\mathrm{r}}[l]]\in\mathbb{C}^{M \times (K+M)}$ and symbol vector $\mathbf{s}[l]\triangleq[\mathbf{s}_{\mathrm{c}}^T[l],\mathbf{s}_{\mathrm{r}}^T[l]]^T$ to simplify the analysis. The received signal at the $k$-th Bob at time slot $t$ is expressed as
\begin{equation}
	\begin{aligned}
		y_k[l]=\mathbf{h}_k^H[l]\mathbf{x}[l]+n_k[l],
	\end{aligned}
\end{equation}
where ${n}_{k}\sim\mathcal{CN}(0,\sigma_k^2)$ stands for the additive white Gaussian noise (AWGN) at  the $k$-th user.
In the multi-user communication system, signals of different users can interfere with one another, and thus the received data rate of Bob $k$ can be obtained as
\begin{equation}
	\begin{aligned}
		R_{\mathrm{b},k}[l]=\log_{2}\left(1+\gamma_{\mathrm{b},k}[l]\right),
	\end{aligned}
\end{equation}
where
\begin{equation}
	\begin{aligned}
		\gamma_{\mathrm{b},k}[l]=\frac{\left|\mathbf{h}_{k}^H[l]\mathbf{w}_{\mathrm{c},k}[l]\right|^2}{\sum_{i\neq k}^K\left|\mathbf{h}_{k}^H[l]\mathbf{w}_{\mathrm{c},i}[l]\right|^2+\sum_{j= 1}^M\left|\mathbf{h}_{k}^H[l]\mathbf{w}_{\mathrm{r},j}[l]\right|^2+\sigma_{k}^2[l]},
	\end{aligned}
\end{equation}

Next, we consider the target sensing. For the Eve, the received signal at time slot $l$ is modeled as
\begin{equation}
	\begin{aligned}
		y_{\mathrm{e}}[l]=\mathbf{h}_{\mathrm{e}}^H[l]\mathbf{x}[l]+n_{e}[l],
	\end{aligned}
\end{equation}
where $n_{\mathrm{e}}\sim\mathcal{CN}(0,\sigma_{\mathrm{e}}^2)$ denotes the AWGN at Eve. We assume that the sensing signal is unknown to the Eve. Correspondingly, the received SINR for the Eve is expressed as
\begin{equation}
	\begin{aligned}
		R_{\mathrm{e},k}[l]=\log_{2}\left(1+\gamma_{\mathrm{e},k}[l]\right),
	\end{aligned}
\end{equation}
where
\begin{equation}
	\begin{aligned}
		\gamma_{\mathrm{e},k}[l]=\frac{\left|\mathbf{h}_{k}^H[l]\mathbf{w}_{\mathrm{c},k}[l]\right|^2}{\sum_{i\neq k}^K\left|\mathbf{h}_{k}^H[l]\mathbf{w}_{\mathrm{c},i}[l]\right|^2+\sum_{j= 1}^M\left|\mathbf{h}_{k}^H[l]\mathbf{w}_{\mathrm{r},j}[l]\right|^2+\sigma_{\mathrm{e}}^2[l]}.
	\end{aligned}
\end{equation}

The signal transmitted to the target is reflected and then received by the BS again, the received echo signal by the BS at time slot $l$ is 
\begin{equation}
	\begin{aligned}
		\mathbf{y}_{\mathrm{r}}[l]=\alpha\mathbf{h}_{\mathrm{e}}[l]\mathbf{h}_{\mathrm{e}}^H[l]\mathbf{x}[l]+\mathbf{n}_{\mathrm{r}}[l],
	\end{aligned}
\end{equation}
where $\mathbf{n}_{\mathrm{r}}\sim\mathcal{CN}(\mathbf{0},\sigma_\mathrm{r}^2\mathbf{I}_N)$ is the AWGN at the BS, and $\alpha$ represents the target radar cross section.
\subsubsection{Secrecy rate (SR)}
In ISAC systems, the SR is a theoretical metric used to quantify the system's ability to ensure communication confidentiality while maintaining sensing performance.  Its core definition typically refers to the information rate that the system can reliably and securely transmit to legitimate users in the presence of eavesdroppers or potentially malicious sensing targets.  The SR for the $k$-th Bob can be expressed as
\begin{equation}
	\begin{aligned}
		R_{\mathrm{sec},k}[l]=[R_{\mathrm{b},k}[l]-R_{\mathrm{e},k}[l]]^+,
	\end{aligned}
\end{equation}
where $(a)^{+}=\operatorname*{max}\{a,0\}$.

\subsubsection{Mutual Information (MI)}
MI is a key metric for measuring the degree of interdependence between two random variables.  In the design of integrated sensing and communication, mutual information is often used to evaluate the correlation between transmitted signals and target echoes.  By calculating the mutual information between transmitted signals and echo signals, it can be determined how much target-related information is contained in the received echoes.  A higher mutual information value indicates stronger correlation between the transmitted signal and the echo, facilitating easier target monitoring and identification.  Conversely, more complex signal processing algorithms may be required to extract target information.  In sensing waveform design, mutual information can be utilized to select optimal sensing waveforms, thereby improving detection performance and target recognition capabilities.  

Specifically, since the BS-Eve channel $\mathbf{h}_{\mathrm{e}}[l]$ contains partial information about the target, the prior uncertainty of the target decreases upon receiving $\mathbf{y}_{\mathrm{r}}[l]$.  Therefore, we employ MI between $\mathbf{h}_{\mathrm{e}}[l]$ and $\mathbf{y}_{\mathrm{r}}[l]$ under the condition of transmitted radar signal $\mathbf{s}_{\mathrm{r}}$, which is denoted as $\mathrm{I}\left(\mathbf{y}_{\mathrm{r}}[l];\mathbf{h}_{\mathrm{e}}[l]\mid\mathbf{s}_{\mathrm{r}}[l]\right)$ to characterize how much information the radar can obtain from $\mathbf{y}_{\mathrm{r}}$,  which is expressed as follows:.
\begin{equation}
	\begin{aligned}
		&\mathrm{I}\left(\mathbf{y}_{\mathrm{r}}[l];\mathbf{h}_{\mathrm{e}}[l]\mid\mathbf{s}_{\mathrm{r}}[l]\right)=h\left(\mathbf{y}_{\mathrm{r}}[l]\left|\mathbf{s}_{\mathrm{r}}[l]\right.\right)-h\left(\mathbf{y}_{\mathrm{r}}[l]\left|\mathbf{h}_{\mathrm{e}}[l],\mathbf{s}_{\mathrm{r}}[l]\right.\right)\\
		&=\frac{1}{2}\log\left(1+\frac{\left|\alpha\right|^{2}\sum_{j=1 }^M\left\|\mathbf{h}_{\mathrm{e}}[l]\mathbf{h}_{\mathrm{e}}^{H}[l]\mathbf{w}_{\mathrm{r},j}[l]\right\|^{2}}{\left|\alpha\right|^{2}\sum_{i= 1}^K\left\|\mathbf{h}_{\mathrm{e}}[l]\mathbf{h}_{\mathrm{e}}^{H}[l]\mathbf{w}_{\mathrm{c},k}[l]\right\|^{2}+\sigma_{{\mathrm{r}}}^{2}[l]}\right).
	\end{aligned}
\end{equation}

\subsection{Problem  Formulation}
	In this paper, we aim to fully leverage the performance advantages offered by RA arrays to enhance both secure transmission and sensing performance. It is noteworthy that, as the multi-user secure transmission performance of the system is constrained by the lower bound, we focus on optimizing the worst user's SR to enhance system robustness. To this end, the objective of our problem is to  maximize the minimum SR for users by jointly optimizing the transmit beamforming matrix $\mathbf{W}[l]$, transmit and receive RAs' pointing matrix $\mathbf{F}[l]$ and $\mathbf{T}[l]$. In addition, the decision variable
set is denoted as $\mathcal{X}\triangleq\{\mathbf{W}[l],\mathbf{F}[l],\mathbf{T}[l]\}$. Mathematically, the related optimization problem is formulated as
\begin{subequations}\label{Problem Formulation1}
	\begin{alignat}{2}
		&\quad\max_{\mathcal{X}}\quad\operatorname*{min}_{k\in \mathcal{K}}~~~\frac{1}{L}\sum_{l=1}^{L}R_{\mathrm{sec},k}[l]\label{PF1_1}\\
		&~~~~\mathrm{s.t.}~~~~\gamma_{b,k}[l]\geq \Gamma, \forall k \in \mathcal{K} \label{PF1_2}\\
		&~~~~~~~~~~~\mathrm{I}\left(\mathbf{y}_{\mathrm{r}}[l];\mathbf{h}_{\mathrm{e}}[l]\mid\mathbf{s}_{\mathrm{r}}[l]\right)\geq  \xi,\label{PF1_3}\\
		&~~~~~~~~~~~\|\mathbf{W}[l]\|_F^2\leq P,\label{PF1_4}\\
		&~~~~~~~~~~~0\leq\operatorname{arccos}({\mathbf{f}}_{m}^T[l]\mathbf{e}_{3})\leq\theta_{\mathrm{max}},\forall m \label{PF1_5}\\
		&~~~~~~~~~~~0\leq\operatorname{arccos}({\mathbf{t}}_{n}^T[l]\mathbf{e}_{3})\leq\theta_{\mathrm{max}},\forall n \label{PF1_6}\\
		&~~~~~~~~~~~\|\mathbf{f}_{m}[l]\|=1,\forall m \label{PF1_7}\\
		&~~~~~~~~~~~\|\mathbf{t}_{n}[l]\|=1,\forall n \label{PF1_8}
	\end{alignat}
\end{subequations}
where $\mathbf{F}[l]\triangleq[\mathbf{f}_{1}[l],\mathbf{f}_{2}[l],\ldots,\mathbf{f}_{M}[l]]\in\mathbb{R}^{3\times M}$ and $\mathbf{T}[l]\triangleq[\mathbf{t}_{1}[l],\mathbf{t}_{2}[l],\ldots,\mathbf{t}_{N}[l]]\in\mathbb{R}^{3\times N}$ denote the transmit and receive RAs' pointing matrix, respectively. Constraint \eqref{PF1_1} and \eqref{PF1_2} indicate the communication quality requirements and sensing MI constraint, respectively. Constraint \eqref{PF1_4} indicates transmit power budget.
Constraint \eqref{PF1_5} and \eqref{PF1_6} ensure that the rotation angles of RAs beam does not exceed a specified range, while constraint \eqref{PF1_7} and \eqref{PF1_8} ensure that $\mathbf{f}_{m}$ and $\mathbf{t}_{n}$ are unit vectors.

\section{Proposed MAPPO-T Algorithm-based Secure Transmission Scheme for RA-enable ISAC Systems}
\label{MAPPO}
Regarding the joint optimization problem of transmit beamforming, pointing matrices of RAs  in low-altitude ISAC systems,  problem \eqref{Problem Formulation1} naturally possesses the characteristics of multi-agent collaborative decision-making. However, traditional convex optimization methods are constrained by the non-convexity of high-dimensional variables and the timeliness deficiency in dynamic user scenarios to solve the problem.  Moreover, single-agent reinforcement learning suffers from the explosion of action space dimensions and the lack of collaborative logic, making it difficult to fully exploit the beam steering DoF of multiple RA units.  Therefore, by adopting the centralized training with decentralized execution framework of the MAPPO algorithm, each RA unit is treated as an independent agent, using the global max-min secrecy rate as the reward to guide cooperative policy learning \cite{Yu2024}.  Trust region constraints are employed to avoid multi-agent policy oscillation, effectively reducing the complexity of the high-dimensional action space, and enabling real-time policy updates in dynamic user scenarios.


\subsection{Markov Decision Process}
The RAs' pointing decision problem can be formulated as a discrete-time sequential decision-making problem. To address the multi-objective optimization problem of time sequential decision planning, the problem is mapped to a decentralized partially observable Markov Decision Process. In the system, all users acts as intelligent agents responsible for decision-making and regulating the system's behavior. Thus, a multi-agent partially observable Markov Decision Process is defined by the tuple $(\mathcal{S}, \mathcal{O}, \mathcal{A}, \mathcal{R}, \mathcal{P})$, which includes the state space $\mathcal{S}$, local observation set $\mathcal{O}$, action space $\mathcal{A}$, reward function $\mathcal{R}$ for each agent, and state transition $P$. These elements are crucial for accurately modeling and solving the Markov decision process, as detailed below.

\begin{itemize}
	\item{}State space $\mathcal{S}$: The state of the system is a description of all users and UAV status,  including channel information for $k$-th user  $\{\mathbf{h}^{(n)}_{i}\}$ and UAV $\{\mathbf{h}^{(n)}_{\mathrm{e}}\}$, as well as their corresponding achievable rates $\{R^{(n)}_{\mathrm{b},k}[l]\}$ and $\{R^{(n)}_{\mathrm{e},k}[l]\}$, respectively. Thus, the state of the system at time slot $n$ is defined as
	\begin{equation}
		{s}^{n}=\{\{R^{(n)}_{\mathrm{b},k}[l]\},\{R^{(n)}_{\mathrm{e},k}[l]\}, \{\mathbf{h}^{(n)}_{k}\},  \{\mathbf{h}^{(n)}_{\mathrm{e}}\}\}, 
	\end{equation}
	where $\boldsymbol{s}^n \in \mathcal{S}$, $\mathcal{S}$ is the set of all state space.
	
	\item{}Observation set $\mathcal{O}$: The information is observed by agent  is the real time zenith and azimuth angles of the pointing vector. Thus, the observations of $i$-th agent at time slot $n$ can be expressed as
	\begin{equation}
		\begin{aligned}
		&{o}_i^{n}=\\
		&\{\sin\theta^{(n)}_{\mathrm{z},i}[l]\cos\theta^{(n)}_{\mathrm{a},i}[l],\sin\theta^{(n)}_{\mathrm{z},i}[l]\sin\theta^{(n)}_{\mathrm{a},i}[l],\cos\theta^{(n)}_{\mathrm{z},i}[l]\},
	\end{aligned}
	\end{equation}	 	
	where $\boldsymbol{o}_i^{n} \in \mathcal{O}$, $\mathcal{O}$ is the set of all observation.
	
	\item{}Action space $\mathcal{A}$: The actions of the intelligent agents are consistent with the optimization variables of the system, thus including the transmit beamforming matrix $\mathbf{W}^{(n)}$, the transmit and receive RAs' pointing matrices  $\mathbf{F}^{(n)}$ and $\mathbf{T}^{(n)}$ of the BS. The action of the system at time slot $n$ is defined as
	\begin{equation}
		a^{n}=\{\mathbf{W}^{(n)}[l],\mathbf{F}^{(n)}[l],\mathbf{T}^{(n)}[l]\},
	\end{equation}	 	
	where  $\boldsymbol{a}_n \in \mathcal{A}$, $\mathcal{A}$ is the set of all action set.
	
	
	\item{}Reward function $\mathcal{R}$: In this work, we have considered MAPPO with a common reward function. The reward function at time slot $n$ is defined by ${r}^n$, which is based on the SR achieved by all users from the environment, and can be calculated as
	\begin{equation}
		r^n= \sum_{l=1}^{L}R^{(n)}_{\mathrm{sec},k}[l].
	\end{equation}

	\item{}State transition $P$: Transitioning from observation state ${o}_i^{n}$ to state ${o}_i^{n+1}$ requires the agent $i$ to complete the switch from action ${a}^{n}$ to action ${a}^{n+1}$.
	
\end{itemize}


\begin{algorithm}[t]
	\caption{Prpoposed MAPPO-T Algorithm for Solving \eqref{Problem Formulation1}}
	\label{MAPPO Algorithm}
	\begin{algorithmic}[1]
		\STATE Initialize policy network parameters $\theta_0$, value function parameters $\phi_0$
		\FOR{$p = 1, 2, \dots, Episodes$}
		\STATE Create empty buffer $D$ for experience replay
		\FOR{$q = 1, 2, \dots, Batch size$}
		\STATE Initialize empty trajectory list $\tau$
		\FOR{$n = 1, 2, \dots, N$}
		\FOR{each agent $i$}
		\STATE Select action based on policy network and current observation ${a}_{i}^{n} = \pi_{\theta_p}({o}_{i}^{n})$;
		\STATE Compute value of current state using value function: $v_{i}^n = V_{\phi_p}(s_{i}^n)$;
		\STATE  Sample a random noise vector $\mathbf{x}_{i}^{n} \sim \mathcal{N}(0, \sigma^2)$ for each agent;
		\STATE  Compute the noisy value $v_{i}^{n}(\phi)$ according to Eq. \eqref{noise};
		\ENDFOR
		\STATE Execute all agents' actions ${a}^{n}$, observe environment feedback $o^{n}_{i}$, rewards ${r}^{n}$, next state ${s}^{n+1}$, next observations $o^{n+1}_{i}$;
		\STATE Add current time step data to trajectory $\tau = [\tau; s^{n}_{i}, o^{n}_{i}, a^{n}, r^{n}, s^{n+1}_{i}, o^{n+1}_{i}]$;
		\STATE Compute expected reward ${R}_{i}^{n}$ according to Eq. \eqref{reward};
		\STATE Compute advantage estimate $\hat{A}_{i}^{n}$ according to Eq. \eqref{GAE};
		\ENDFOR
		\STATE Merge trajectory $\tau$ into buffer $D$;
		\ENDFOR
		\STATE Update policy network parameters $\theta_p \to \theta_{p+1}$ using Adam optimizer according to Eq. \eqref{policy} ;
		\STATE Update value network parameters $\phi_p \to \phi_{p+1}$ using Adam optimizer according to Eq. \eqref{value};
		\ENDFOR
	\end{algorithmic}
\end{algorithm}

\subsection{Standard MAPPO algorithm}
\begin{figure}[!t]
	\centering
	\includegraphics[width=0.44\textwidth]{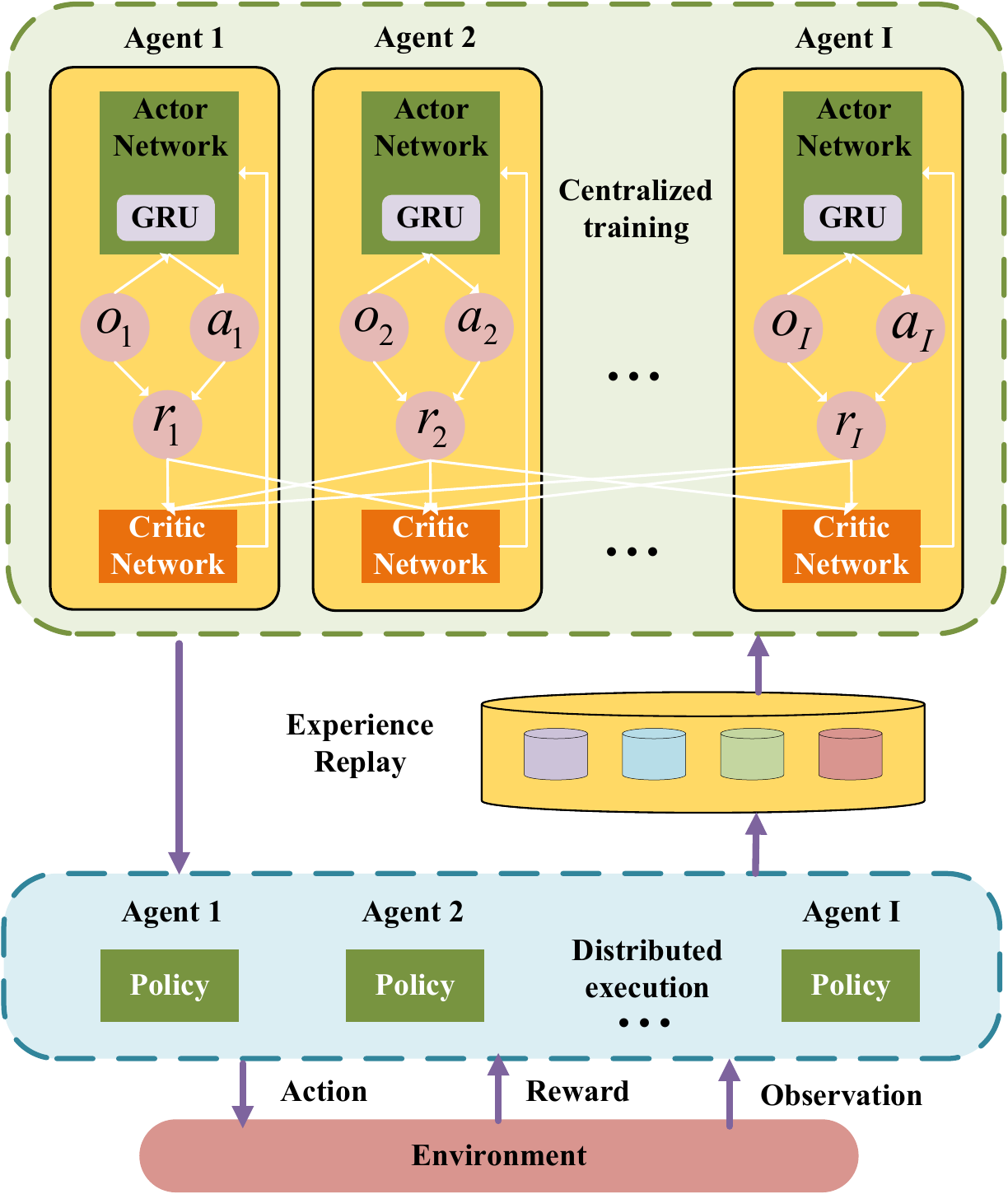}\\
	\caption{Framework of the MAPPO-T algorithm.}\label{Fig1.eps}
\end{figure}

The MAPPO algorithm is a reinforcement learning method that extends the single-agent proximal policy optimization algorithm to multi-agent environments, typically employing a centralized training and distributed execution framework. Under this framework, each agent possesses its own policy network and value network; however, during training, the value function can be updated using global state information and the joint actions of all agents, thereby mitigating the issue of environmental non-stationarity. The flowchart of the MAPPO-T method is shown in Algorithm \ref{Fig1.eps}.

The number of agents is set to $I=M+N$, for each agent $i$, the policy $\pi_{\theta_i}(a_i|o_i)$ is represented by parameters $ \theta_i$, where $o_i$ denotes the local observation of the agent. The global state $s$ and the joint actions $a$ of all agents are taken as inputs to the value function $V_{\phi_i}(s, a)$, by which the joint state-action value is estimated. For the improvement of training efficiency and stability, the policy network parameters are shared among all agents, whereas neither exploration noise nor independent updates are shared. The target network or the old policy parameters $\theta_{\text{old},i}$ are initialized to be identical to $\theta_i$.

The algorithm first executes the interactive sampling process, deploying the current policy in the environment to collect trajectories. At time step $n$, the global state is $s^n$. Each agent $i$ samples an action $a^n_{i}$ based on its local observation $o_{i}$ using the policy network $\pi_{\theta_i}(a_i|o_i)$, After the joint action ${a}^n = (a_1^{n}, a_2^{n}, \dots, a_I^{n})$ is applied to the environment, the global reward $r^n$ is obtained, and the system transitions to the next state $s^{n+1}$. The trajectory dataset stores local observations, actions, global rewards, global states, and the action probabilities of the old policy before the update, $\pi_{\text{old},i}(a_{i}^n|o_{i}^n)$, providing the foundation for subsequent importance sampling calculations.

For each agent $i$, an estimate of the advantage function is first computed. The generalized advantage estimation (GAE) is introduced to calculate the advantage function to balance estimation bias and variance, the temporal differential residuals for agent $i$ is defined by
\begin{equation}
	\delta_{i}^n=r^n+\gamma V_\psi({s}^{n+1})-V_\psi({s}^n),
\end{equation}
where $V_ \psi (s)$ is the global state value output by the centralized value network, and the GAE function can be expressed as
\begin{equation}\label{GAE}
	\hat{A}_{i}^n =\sum_ {l=0} ^ {N-n-1} (\gamma \lambda) ^ l \delta_{i}^{n+l},
\end{equation}
where $\lambda$ is the attenuation coefficient of the dominance estimation. The centralized value network takes the global state as input, which can effectively characterize the long-term benefits of multi-agent joint actions and improve the estimation accuracy of the dominance function.

The objective function of the actor network $\pi_\theta$ is expressed as the maximization of the expected value of the total expected return, i.e., the maximization of the expectation of the total reward, which is expressed as 
\begin{equation}\label{policy}
	\begin{aligned}
		L_i^{\mathrm{CLIP}}(\theta_i)=\mathbb{E}^n\left[\min\left(\rho_{i}^n, \hat{A}_{i}^n,\mathrm{clip}(\rho_{i}^n,1-\varepsilon,1+\varepsilon)\hat{A}_{i}^n\right)\right],
	\end{aligned}
\end{equation}
where $\rho_{i}^n=\frac{\pi_{\theta_i}({a}_{i}^n|{o}_{i}^n)}{\pi_{\text{old},i}({a}_{i}^n|{o}_{i}^n)}$ is the importance sampling ratio, which characterizes difference in the probabilities of the corresponding actions under the old and new policies, thereby preventing the training from diverging due to excessive policy updates. $\varepsilon$ is the shear coefficient, which limits the range of policy updates through truncation operations. When the dominance function is positive, the constraint ratio does not exceed $1+\varepsilon$, and when the dominance function is negative, the constraint ratio is not less than $1- \varepsilon$, ensuring the stability of policy updates.

The overall strategy loss is the weighted sum of all agents cutting the target and the entropy regularization term, which is represented as
\begin{equation}
	H(\pi_{\theta_i})=-\mathbb{E}_{{a}_i\sim\pi_{\theta_i}}[\log\pi_{\theta_i}({a}_i|{o}_i)].
\end{equation}

Therefore, the total strategy objective function can be expressed as
\begin{equation}
	L_{\mathrm{POL}}(\boldsymbol{\theta})=\sum_{i=1}^{N}\left[L_{i}^{\mathrm{CLIP}}(\theta_{i})+c_{2}H(\pi_{\theta_{i}})\right],
\end{equation}
where $\boldsymbol{\theta}=\{\theta_1,\theta_2,\dots,\theta_I\}$ denotes the set of policy parameters for all agents, and $c_2$ is the entropy weight coefficient.

Introducing policy entropy can enhance the ability to explore strategies and avoid premature convergence to local optimal policies, 
The centralized value network updates by minimizing the mean square error, and the target value is determined by the cumulative discount return, which is given by 
\begin{equation}\label{reward}
	R^n=\sum_{l=0}^{\infty}\gamma^l r^{n+l},
\end{equation}
where $\gamma$ represents the discount factor, and $r^{n+l}$ is the local reward. Thus, the value loss function is
\begin{equation}\label{value}
	L_i^{\mathrm{VF}}(\phi_i)=\mathbb{E}^n\left[\left(V_{\phi_i}(s^n,a^n)-\hat{R}^n\right)^2\right].
\end{equation}

\subsection{Three improved mechanisms}
By combining the optimization techniques of the PPO algorithm with multi-agent cooperative systems, the MAPPO algorithm can effectively address collaborative decision-making problems in multi-agent environments. However, the MAPPO algorithm also faces several potential challenges, such as policy overfitting, the coordination of transmission and reception among multiple rotating array units, temporal correlation issues in RAs' pointing decision, and training instability resulting from multi-agent collaborative decision-making, respectively. Therefore, to address these challenges, this paper introduces three improvement mechanisms, i.e., a noise-based processing approach, a recurrent neural network (RNN)-based improvement method, and the incorporation of PopArt technology, respectively.

\subsubsection{Noise-based processing}
The MAPPO algorithm uses a shared advantage value to learn policies for all agents, which may lead to overfitting of policies unrelated to that advantage. This issue is known as Policy Overfitting in Multi-agent Cooperation (POMAC). Due to limited sampling, POMAC can cause some policies to update in a suboptimal direction, thereby preventing the exploration of trajectories with higher returns. This issue may have a particularly significant impact when multiple agents are performing collaborative tasks. To address the potential POMAC problem, noise-based methods can be employed to resolve this issue.

For the $i$-th agent, a Gaussian noise vector ${\mathbf{x}}_{i} \sim  N\left( {0,{\sigma }^{2}}\right)$ is randomly sampled, where ${\sigma }^{2}$ is the variance, representing the noise intensity. 
Combine the noise $\mathbf{x}_{i}$ with the state $S$ and feed the integrated feature into the centralized value network to output an individual value ${v}_{i}$ for each agent.
\begin{equation}\label{noise}
	{v}_{i} = V\left( {\operatorname{concat}\left( {s_i,{\mathbf{x}}_{i}}\right) }\right) ,\forall i = 1,2,\ldots ,K,
\end{equation}

The noise value ${v}_{i}$ is propagated to the advantage value ${A}_{i} = r + \gamma {v}_{i}\left( {s}^{n}\right)  - {v}_{i}\left( {s}^{n + 1}\right)$, thereby smoothing the advantage value.

\subsubsection{GRU-based improvement}
In the standard MAPPO algorithm, multilayer perceptrons are adopted by default to construct the Actor policy network and the centralized Critic value network. Between network layers, only single-layer feature mapping and nonlinear transformation are performed. Consequently, instantaneous decision-making modeling is enabled solely based on local observations and global states at a single time step. The capability to capture temporally dependent information is lacking, rendering the standard MAPPO algorithm poorly adaptable to the continuous temporal coupling characteristics inherent in multi-user cooperative communication and dynamic trajectory planning scenarios.

To address this issue, an RNN module replaces the original feedforward structure in MAPPO, leveraging temporal memory and state backtracking to encode historical observations, motion trajectories, and time-varying communication environments. When the critic and policy networks are modified from multilayer perceptrons to RNNs, the loss function is summed over time, and the networks are trained via backpropagation through time. However, when long sequences are processed, potential vanishing or exploding gradient problems are inherent to the RNN, by which the learning of long-term dependencies is rendered difficult for the network. To address this problem, the Gated Recurrent Unit (GRU) was proposed to control the flow of information and effectively alleviate the problems of gradient vanishing and exploding by introducing gating mechanisms, which enabling the network to better capture long-term dependencies, improving the performance and generalization ability of the model. GRU introduces two most important gating units: reset gate and update gate, which are updated as follows:
\begin{equation}
	R^n=\sigma(X_tW_{xr}+cW_{hr}+b_r),
\end{equation}
\begin{equation}
	Z^n=\sigma(X_tW{xz}+H^{n-1}W_{hz}+b_z),
\end{equation}
where  $X^n$  is a small batch input sample and $H^{n-1}$  is the hidden state of the previous time step. $W_{xr}$ and $W{xz}$ are weight parameters, $b_r$ and $b_z$ are bias parameters, respectively. Integrating the reset gate with the conventional hidden state update mechanism can obtain candidate hidden states at the time step $n$:
\begin{equation}
	\tilde{H}^n=\tanh(X_tW_{xh}+(R^n\odot H^{n-1})W_{hh}+b_h),
\end{equation}
where $W_{xh}$ and  $W_{hh}$ are weight parameters, $b_h$ is bias parameter. The hidden state of the current time step can be calculated by the following equation:
\begin{equation}
	H^n=Z_t\odot H^{n-1}+(1-Z_t)\odot\tilde{H}^n.
\end{equation}

\subsubsection{Incorporation of PopArt}
Due to the variability and randomness of multi-agent environments, the magnitude of rewards may vary significantly across different environments or time steps, which can lead to instability in the training process. In particular, in multi-agent environments, the reward distributions of different agents may differ significantly. Therefore, the MAPPO algorithm incorporates the PopArt  technique to address the reward normalization problem in multi-agent environments and achieve stable training. The core idea of PopArt is to maintain the stability of the output by adaptively rescaling the weights and biases of the neural network’s output layer, while simultaneously rescaling the targets. Specifically, PopArt is capable of adaptively normalizing and rescaling the targets without altering the already learned outputs.

Firstly, during the training process, PopArt requires calculating the mean and variance of rewards to maintain the moving average and variance of the rewards. The exponential weighted moving average method can achieve the calculation according to 
\begin{equation}
	{\mu}^{n} = \left( {1 - \beta}\right) {\mu}^{n - 1} + \beta {R}^{n}, 
\end{equation}
\begin{equation}
	({\sigma }^{n})^2 = \left( {1 - \beta }\right) ({\sigma}^{n-1})^{2} + \beta {\left( {R}^{n} - {\mu}^{n}\right) }^{2}, 
\end{equation}
where $ {R}^{n} $ is the current reward, $ {\mu}^{n} $ and $ {\sigma }^{2} $ are the mean and variance of the reward, respectively, and $ \beta $ is a small learning rate.

Next, the reward values are standardized using the statistical mean and variance.
\begin{equation}
	{R}^{n^{\prime}}= \frac{{R}^{n} - {\mu }^{n}}{{\sigma }^{n}}.
\end{equation}

Finally, the output layer of the neural network is adjusted, and the weights and biases of the output layer are re-adjusted so that the network output remains unchanged. Specifically, the output layer of the neural network is assumed to be a linear layer $ y = {Wz} + b $, where $ W $ is the weight matrix, $ b $ is the bias vector, and $ z $ is the output of the hidden layer. 

In order to keep the output unchanged, the re-adjusted weights and biases are given as follow
\begin{equation}
	{W}^{\prime } = \frac{W}{{\sigma }^{n}},  {b}^{\prime } = b - \frac{W{\mu }^{n}}{{\sigma }^{n}}
\end{equation}

\subsection{Computational Complexity Analysis}
To evaluate the efficiency of the proposed solution, we mainly analyze the computational complexity of the Algorithm \ref{MAPPO Algorithm}. Assuming a multi-agent system consists of $I$ agents, with a single environmental interaction trajectory length of $T$, the dimensions of the single-layer hidden layer neurons in the policy network and centralized value network are unified as $E$, and the total number of network layers are $L_\pi$ and $L_v$, respectively. The number of batch divisions in a single sampling round of the algorithm is $B$, the number of rounds of parameter iteration updates is $P$, the policy output dimension and the single agent action space dimension in continuous action scenarios are denoted as $d_a$, the local observation dimension of the agent is $d_o$, and the global state input dimension is $d_s$. In the environmental trajectory sampling stage, each agent independently completes policy inference based on local observations. The basic computational complexity of forward propagation in a single agent policy network is determined by the input dimension, network layers, and neuron size. The complexity of a single forward inference can be expressed as $O(L_\pi d_o E+L_\pi E d_a)$. The overall inference complexity of $K$ agents completing action sampling in parallel is $O(P(L_\pi d_o E+L_\pi E d_a))$. The centralized value network takes the global state as input, and the complexity of single step state value inference is $O(L_v d_s E)$. In the sampling process with a complete length of $T$, the total time complexity of this stage is $O\left[T\left(I L_\pi E(d_o+d_a)+L_v E d_s\right)\right]$, and the sampling process only involves forward matrix operations and nonlinear activation calculations.

The core computing overhead is concentrated in the parameter optimization stage of multiple batches and rounds. The proposed algorithm adopts the goal of cutting PPO to complete single agent policy updates, and all agents share centralized value network parameters and independently maintain their own policy parameters. The sample size for a single batch is $IT/B$, and the complexity of solving the policy loss and backpropagation for a single agent in a single iteration is $O(L_\pi E(d_o+d_a)IT/B)$. The complexity of calculating the policy gradient after stacking $I$ agents is $O\left({I L_\pi E(d_o+d_a)T}/{B}\right)$; The centralized value network completes value fitting and gradient updating for global samples, with a backpropagation complexity of $O\left({L_v E d_s T}/{B}\right)$. Combined with the algorithm's built-in $P$-round iterative optimization mechanism, the total time complexity of the complete parameter update phase can be integrated as $O\left({PT}\Big(I L_\pi E(d_o+d_a)+L_v E d_s\Big)/{B}\right)$.

\section{Simulation Results}
\label{Simulation Results}
\begin{table}[!t]
	\centering
	\caption{Simulation Parameters}
	\label{parameters}
	\begin{tabular}{|c|c|}
		\hline
		\textbf{Parameters} & \textbf{Value} \\
		\hline
		Position variable of the UPA center & \( d_0 = 2 \, \text{m} \) \\
		\hline
		Number of antennas at the BS    & \( M = M = 9 \) \\
		\hline
		Wavelength                      & \( \lambda = 0.01 \, \text{m} \) \\
		\hline
         Antenna separation           & \(d={\lambda}/{2} \) \\
		\hline
		Number of scatterer clutters              & \( Q = 6 \)   \\
		\hline
		Number of users                 & \( K = 3 \)   \\
		\hline
		 Initial distance between RAs to users &  $r_1 = r_2 =r_3 = 50$ m   \\
		 \hline
		 \begin{tabular}[c]{@{}c@{}}Initial zenith angles of users relative \\ to the origin of the coordinate system\end{tabular} &  \begin{tabular}[c]{@{}c@{}}$\phi_1 = \pi/12$,  $\phi_2= \pi/6$, \\ $\phi_3= \pi/4$ \end{tabular}   \\
		  \hline
		 \begin{tabular}[c]{@{}c@{}}Initial azimuth angles of users  relative\\  to the origin of the coordinate system\end{tabular} &  \begin{tabular}[c]{@{}c@{}}$\psi_1 = -\pi/12 $,  $\psi_2= 0$, \\  $\psi_3= \pi/12$ \end{tabular}   \\
		\hline
		Velocity of users               & \begin{tabular}[c]{@{}c@{}} $v_{k,x}=v_{k,y}$ = 1 m/s, \\ $v_{k,z}$= 0 m/s \end{tabular} \\
		\hline
		Initial distance between RAs to UAV &  $r_0 = 60$ m   \\
		\hline
		\begin{tabular}[c]{@{}c@{}}Initial zenith angle of UAV  relative \\ to the origin of the coordinate system\end{tabular} &  	 $\phi_{\mathrm{e}}= 2\pi/3$    \\
		\hline
		\begin{tabular}[c]{@{}c@{}}Initial azimuth angle of UAV  relative \\ to the origin of the coordinate system\end{tabular} &   $\psi_{\mathrm{e}}= \pi/12$  \\
		\hline
		Velocity of UAV               &  $v_{e,x}= v_{e,y}=v_{e,z}$ = 3 m/s \\
		\hline
		BS power budget                    & \( P_{\mathrm{BS}} = 30 \, \text{dBm} \) \\
		\hline
		Noise power for users \& target & \( \sigma_k^2 = \sigma_{\mathrm{e}}^2 = -80 \, \text{dBm} \) \\
		\hline
	\end{tabular}
\end{table}

\begin{table}[!t]
	\centering
	\caption{Neural Network Structure of MAPPO-T.}
	\label{Network Structure}
	\begin{tabular}{|c|c|c|c|}
		\hline
		Policy & Input & Output & Activation \\
		\hline
		1st FC & (None, norm(\#obs)) & (None, 128) & relu $\rightarrow$ norm \\
		\hline
		2nd FC & (None, 128) & (None, 128) & relu $\rightarrow$ norm \\
		\hline
		3rd FC & (None 128) & (None 128) & relu $\rightarrow$ norm \\
		\hline
		4th FC & (None, 128) & (None, \#acts) & categorical \\
		\hline\hline
		Value Func & Input & Output & Activation \\
		\hline
		1st FC & (None, norm(\#states)) & (None, 128) & relu $\rightarrow$ norm \\
		\hline
		2nd FC & (None, 128) & (None, 128) & relu $\rightarrow$ norm \\
		\hline
		3rd FC & (None 128) & (None 128) & relu $\rightarrow$ norm \\
		\hline
		4th FC & (None, 128) & (None, 1) & none \\
		\hline
	\end{tabular}
\end{table}

In this section, the simulation results evaluate the performance of RA-enabled secure ISAC systems with the proposed MAPPO-T algorithm.
In simulation settings, the BS is the origin of the global coordinate system. The BS operates at a carrier frequency of 2.4 GHz with a wavelength of $\lambda=0.125$ meter (m), which serving $K = 3$ users and sensing one UAV, with the interference of $Q = 6$ scatterer clusters \cite{Zheng2026}.  The center coordinates of the transmitting UPA and the receiving UPA are set to $[0,0,d_0]^T$ and $[0,0,2d_0]^T$, respectively. The BS is equipped with $M$ transmit RAs and $N$ receive RAs.  The RA array is centered at the origin, with an inter-antenna spacing of $d = {\lambda}/2 $. 
We assume that the initial legitimate users' zenith and azimuth angles with respect to the origin of the coordinate system are set as $\phi_1 = \pi/12$,  $\phi_2= \pi/6$ and $\phi_3= \pi/4$,  as well as $\psi_1 = -\pi/12 $,  $\psi_2= 0$ and $\psi_3= \pi/12$ with $r_1 = r_2 =r_3 = 50$ m are the distances between RAs to users, respectively. We set users' velocity $[v_{1,x},v_{1,y}, v_{1,z}]$= [1 m/s, 1  m/s, 0 m/s], $[v_{2,x},v_{2,y}, v_{2,z}]$= [1 m/s, 1  m/s, 0 m/s] and $[v_{3,x},v_{3,y}, v_{3,z}]$= [1 m/s, 1  m/s, 0 m/s], respectively.  Moreover, the initial UAV's zenith and azimuth angles with respect to the origin of the coordinate system are set as  $\phi_{\mathrm{e}}= 2\pi/3$   and $\psi_{\mathrm{e}}= \pi/12$ with velocity $[v_{e,x},v_{e,y},v_{e,z}]$= [3 m/s, 3  m/s, 3 m/s], and  $r_0=60$ m denotes the distance from the origin to UAV.  The average noise power is $\sigma_k^2 = \sigma_{\mathrm{e}}^2 = -80$ dBm.  
Unless otherwise specified, the number of users, the number of transmit and receive RAs, the communication SINR requirement, the transmit power budget of BS,  the MI threshold, and the maximum zenith angle are set to $K=3$, $M = N = 9$, $P_{\mathrm{BS}}$ = 30 dBm, $\Gamma$ = 5 dB, $\xi$ = 1.5 and $\theta_{\mathrm{max}} = \pi/6$, respectively. The simulation parameters for the proposed system are listed in Table \ref{parameters}.

The experiments took place on a Windows platform equipped with an Intel Core i7 processor, a GeForce 4070 graphics card, and 16GB of RAM. The architectures of strategy network and value neural network are detailed in Table \ref{Network Structure}, where FC denotes the fully connected layer. Both neural networks use the Adam optimizer during training, with learning rates set at $5 \times 10^{-4}$ and $10^{-5}$, respectively. The clipping coefficient $\varepsilon$ and strategy entropy coefficient $\sigma$ in formula  are set to 0.2 and 0.01, respectively. The discount factor $\gamma$ and smoothing factor $\lambda$ in Eq. \eqref{GAE} are set to 0.99 and 0.95, respectively. In addition, the time step, the batch size and the number of training epochs is set to 300, 1, 1000, respectively.

To further validate the performance advantages of the proposed MAPPO-T algorithm in RA system, we consider other three benchmark methods for comparison, which are listed as follows:
\begin{itemize}
	\item{}\textbf{MAPPO:} The standard MAPPO algorithm, with parameter settings consistent with those of the MAPPO-T algorithm.
	\item{}\textbf{Fixed orientation antenna (FOA):} In this scheme, the orientations of all RAs are fixed at their reference orientations, i.e., $ \mathbf{f}_n = \mathbf{e}_3, \forall n$,  and the MMSE receive beamforming is applied at the BS.
	\item{}\textbf{Random orientation antenna (ROA):} In this scheme, the orientation of each RA is randomly generated within the
	rotational ranges given by \eqref{jiaodu1} and \eqref{jiaodu2}, and the MMSE receive
	beamforming is applied at the BS.
\end{itemize}

\begin{figure}[!t]
	\centering
	\includegraphics[width=3.1in]{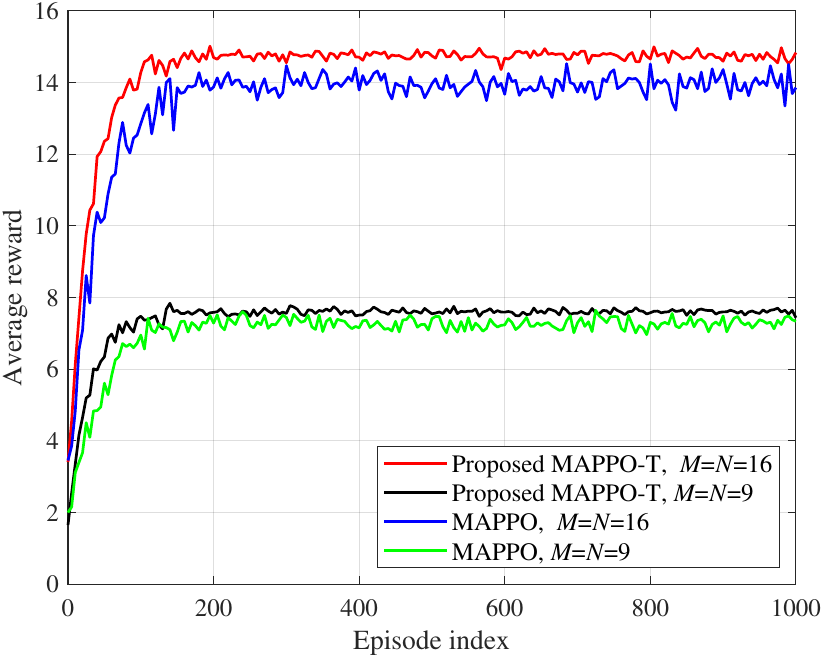}
	\caption{Convergence behavior of the proposed MAPPO-T compared to MAPPO algorithm.}
	\label{convergence}
\end{figure}
Fig.~\ref{convergence} illustrates the convergence behavior of the proposed MAPPO-T method compared to MAPPO benchmark algorithm. The simulation results show that, under the same RA configuration, MAPPO-T demonstrates superior convergence performance compared to the standard MAPPO algorithm across the board. Furthermore, increasing the number of RA units yields significant performance gains. The reason is shown that more RA units provide greater beamforming flexibility, enabling simultaneous enhancement of legitimate user signals, suppression of eavesdropper interference, and improved sensing accuracy. Furthermore, the performance advantage of MAPPO-T in high-dimensional configurations demonstrates its adaptability to high-dimensional optimization spaces, whereas the standard MAPPO algorithm is unable to fully utilize the additional DoF due to constraints related to overfitting and training stability. More importantly, the amplitude of reward fluctuations during the convergence phase is smaller for MAPPO-T than for the standard MAPPO, validating the superior adaptability and robustness of its strategy in dynamic user scenarios.

\begin{figure}[!t]
	\centering
	\includegraphics[width=3.1in]{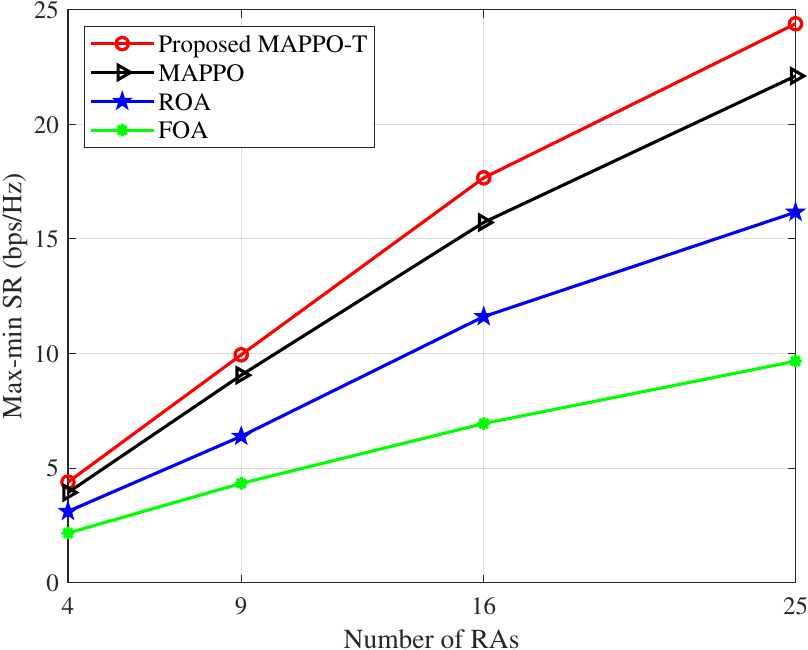}
	\caption{The max-min SR versus the number of RAs $M=N$.}
	\label{N_antenna}
\end{figure}
Fig.~\ref{N_antenna} illustrates the max-min SR behavior of versus the number of RAs $M=N$. As shown in the figure, the max-min SR of all algorithms increases monotonically with the number of RAs. When the number of RAs is 16, the max-min SR of the MAPPO-T algorithm reaches 17.6 bps/Hz. This is attributed to the fact that an increased number of RA units enhances the system’s beam control flexibility, thereby strengthening legitimate user signals while simultaneously suppressing eavesdropper signals, which leads to an improvement in the security rate. Furthermore, the proposed MAPPO-T algorithm consistently outperforms the comparison algorithms. This is because the standard MAPPO algorithm is limited by overfitting and training instability, preventing it from fully utilizing the RA unit’s DoF, resulting in suboptimal performance. Meanwhile, the ROA algorithm employs a random RA pointing strategy, leading to a lack of specificity in beam control and thus limited improvement in the security rate. The FOA algorithm, which uses a fixed RA pointing strategy, cannot adapt to the dynamic channel conditions caused by user mobility, resulting in the worst performance.

\begin{figure}[!t]
	\centering
	\includegraphics[width=3.1in]{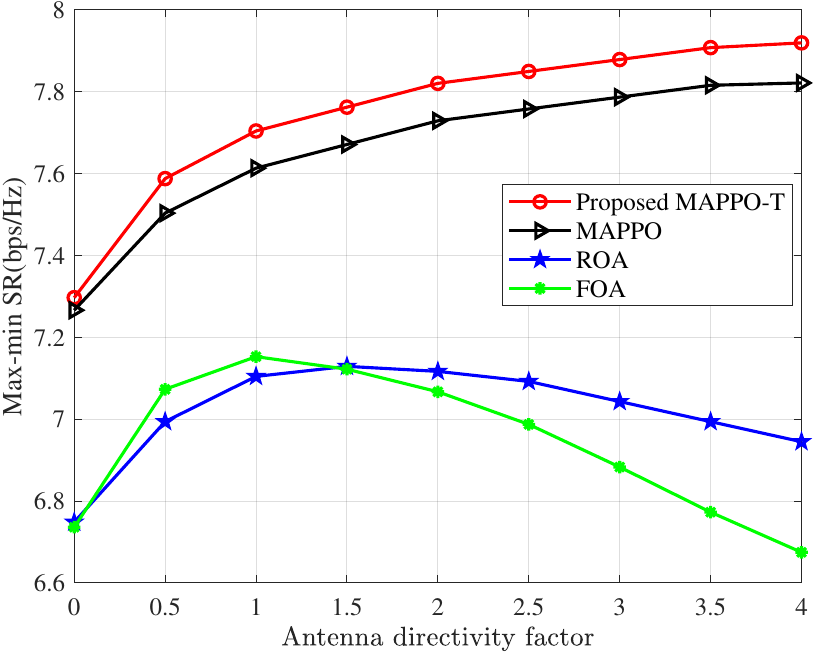}
	\caption{Max-min SR versus the antenna directivity factor $p$.}
	\label{directivity_factor}
\end{figure}
Fig.~\ref{directivity_factor} depicts the max-min SR behavior versus the antenna directivity factor $p$. When $p=4$ , the MAPPO-T algorithm achieves a maximum-minimum secure data rate of approximately 7.9 bps/Hz, representing an improvement of about 1.1\% over the standard MAPPO algorithm. In addition, the max-min SR of MAPPO-T and MAPPO algorithms monotonically increases with the increase of antenna directional factor, while the max-min SR of ROA and FOA algorithms shows a trend of first increasing and then decreasing with the increase of antenna directional factor. This is attributed to the fact that the larger $p$ of the antenna, the narrower the main lobe width of the beam, and the more concentrated the energy is towards the target direction. Therefore, RA systems with larger p-values can more effectively enhance directional gain in the multi-user direction, thereby achieving greater max-min SR. However, ROA and FOA algorithms lack dynamic optimization capabilities and cannot adjust beam strategies with changes in $p$. When $p$ is too large, the beam becomes too narrow, resulting in decreased signal coverage for legitimate users during movement and ineffective interference suppression by eavesdroppers, leading to a decrease in confidentiality rate. Therefore, the proposed MAPPO-T algorithm maintains stable performance advantages in adapting to dynamic environments under different $p$ conditions.

\begin{figure}[!t]
	\centering
	\includegraphics[width=3.1in]{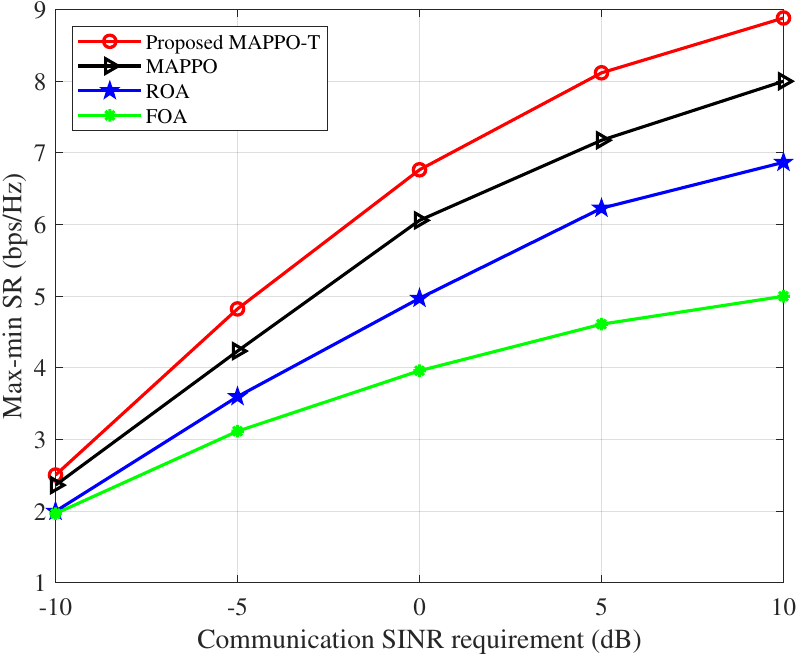}
	\caption{Max-min SR versus the communication SINR requirement $\Gamma$.}
	\label{Gamma}
\end{figure}
Fig.~\ref{Gamma}  describes the the max-min SR versus the communication SINR requirement $\Gamma$. It is evident that the max min secrecy rate of all algorithms monotonically increases with the increase of communication SINR requirements. The proposed MAPPO-T algorithm achieves a performance improvement of about 11.3\% compared to the standard MAPPO algorithm when the communication SINR requirement is 10dB. This is due to the fact that the MAPPO-T algorithm can effectively improve the SR by jointly optimizing  beamforming and RA pointing matrix to enhance legitimate user signals and suppress eavesdropper interference under higher SINR constraints, demonstrating the stable performance advantage and  the robustness of the MAPPO-T algorithm strict communication constraints.

\begin{figure}[!t]
	\centering
	\includegraphics[width=3.1in]{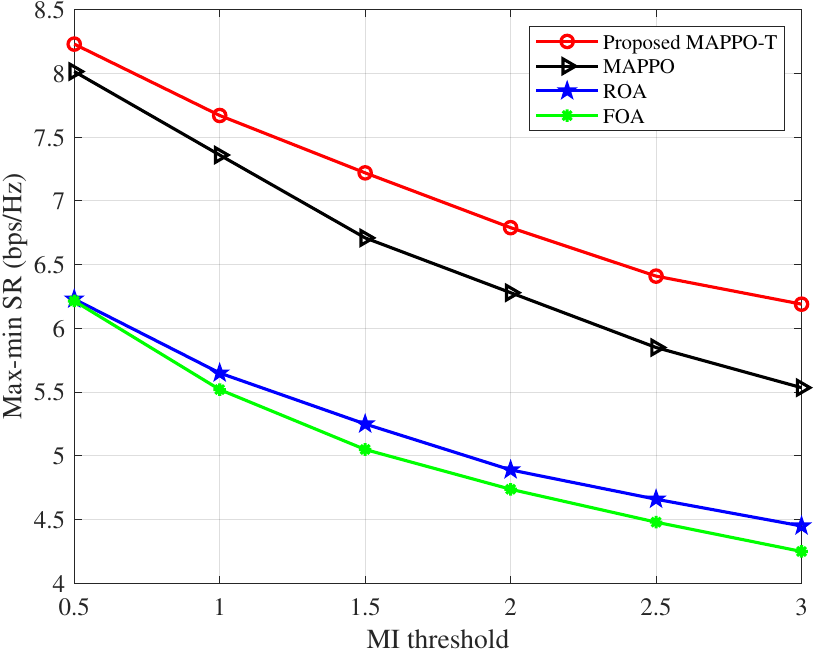}
	\caption{Max-min SR versus the MI threshold $\xi$.}
	\label{MI}
\end{figure}
In Fig.~\ref{MI}, we analyze the relationship between max-min SR and MI threshold $\xi$. We observed that the max-min SR of all algorithms monotonically decreases with increasing mutual information threshold, and the performance of the proposed MAPPO-T algorithm consistently outperforms other compared algorithms. This is attributed to the fact that the increase in mutual information threshold imposes higher constraints on the target perception performance, and the system needs to allocate more resources to meet the perception requirements, resulting in a reduction in the resources available on the communication side to improve the confidentiality rate. The MAPPO-T algorithm can dynamically balance communication and perception resources by jointly optimizing base station beamforming and RA pointing matrix, and maintain high confidentiality rates even under high mutual information thresholds.


\begin{figure}[!t]
	\centering
	\includegraphics[width=3.1in]{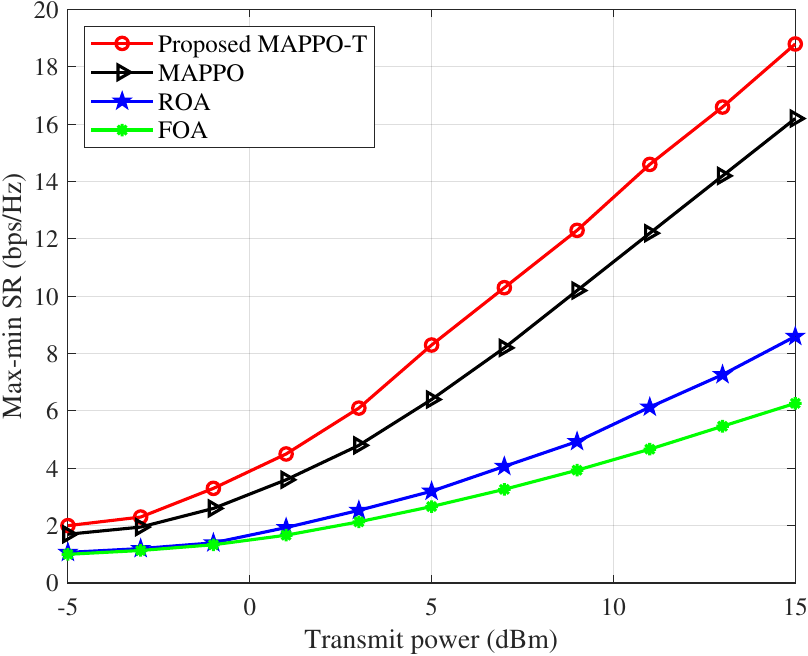}
	\caption{Max-min SR  versus the transmit power  $P_{\mathrm{BS}}$.}
	\label{Tramsmmit_P}
\end{figure} 
Fig.~\ref{Tramsmmit_P}   illustrates the max-min SR  behavior comparison against the transmit power $P_{\mathrm{BS}}$. We can observe a clear trend that the max-min SR of all algorithms monotonically increases with the transmit power $P_{\mathrm{BS}}$.  This is because the increase in $P_{\mathrm{BS}}$ provides higher signal DoF, enhances the signal strength of legitimate users, and suppresses eavesdropper interference, thereby improving the secrecy rate.  Moreover, the MAPPO-T algorithm fully exploits the additional power resources to maximize the improvement in secrecy performance, while the ROA and FOA algorithms lack and effective utilization, resulting in limited performance improvement.  Therefore, the performance advantage of the MAPPO-T algorithm under high transmit power becomes increasingly significant as power increases, reflecting its optimization capability and robustness in scenarios with abundant power resources.

\begin{figure}[!t]
	\centering
	\includegraphics[width=3.1in]{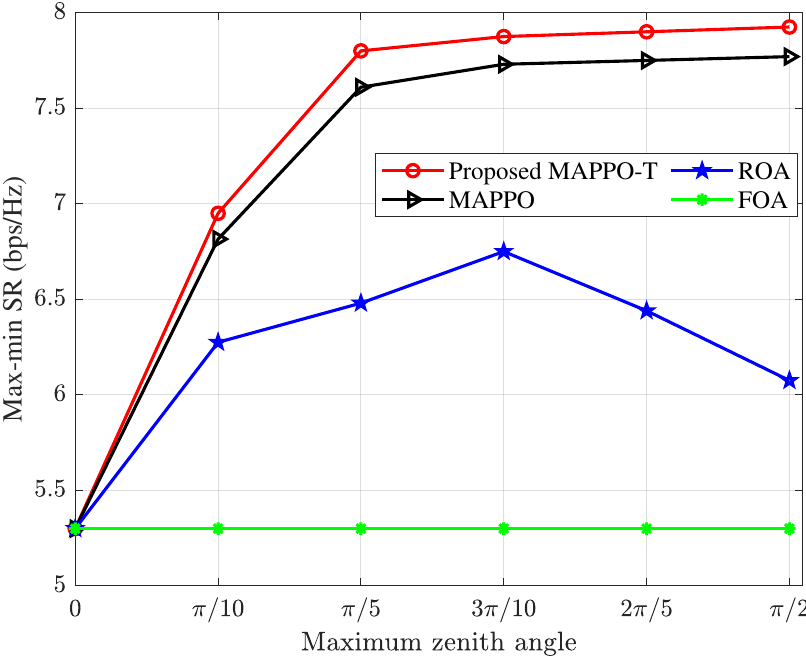}
	\caption{Max-min SR versus the maximum zenith angle $\theta_{\mathrm{max}}$.}
	\label{zenith_angle}
\end{figure} 
Finally, Fig.~\ref{zenith_angle}  demonstrates the max-min SR versus the maximum zenith angle $\theta_{\mathrm{max}}$.  First, as$\theta_{\mathrm{max}}$ increases, the max-min secrecy rate of the proposed MAPPO-T algorithm first increases rapidly and then stabilizes.  Specifically, when $\theta_{\mathrm{max}}\leq \pi/10$, the performance improvement is significant, and a small rotation adjustment range can achieve obvious performance gains.  When $\theta_{\mathrm{max}}$ exceeds $\pi/5$, the performance tends to stabilize, and the RA system obtains sufficient DoF to balance the array directional gains in a multipath channel.  Furthermore, the proposed RA system always outperforms the ROA and FOA systems, because the latter two systems cannot independently adjust the direction of each individual antenna to reconfigure the directional gain pattern of the entire array.  The random directional gain of the ROA system allows it to radiate signals in any direction, thus achieving a better rate gain than the FOA system.  However, the RA pointing of the ROA system cannot form an ordered array directional gain pattern as the zenith angle expands, resulting in performance loss.

%

\section{Conclusion}
\label{Conclusion}
In this paper, we studied a novel RA-enabled active RIS-aided ISAC systems to enhance the secure transmission.   
The core objective of our work is to maximize the minimum SR of legitimate mobile users, which is achieved by jointly optimizing  the transmit beamforming matrix, the pointing matrices of RA arrays.
To address the complex multi-agent collaborative decision-making problem, we proposed a multi-agent proximal policy optimization algorithm with three improvement mechanisms, termed MAPPO-T. 
Simulation results demonstrated that the introduction of RAs significantly improves SR performance compared to conventional FOA-based systems. Furthermore, the proposed MAPPO-T algorithm exhibits superior performance over the standard MAPPO algorithm, validating the effectiveness of the three incorporated improvement mechanisms.

\renewcommand\refname{References}
\bibliographystyle{IEEEtran}
\bibliography{manuscript}
\end{document}